\documentclass[11pt,letterpaper]{softmax_preprint}

\usepackage{multirow}
\usepackage{textcomp}
\usepackage{tabularx}

\setcitestyle{numbers,square,comma}
\newtcolorbox{rqsummarybox}{%
  enhanced, breakable,
  colback=abstractbg, colframe=abstractbg,
  boxrule=0pt, arc=3pt, outer arc=3pt,
  left=8pt, right=8pt, top=4pt, bottom=4pt,
  fontupper=\small, before skip=10pt, after skip=10pt
}
\newcommand{\rqsummarylabel}[1]{%
  \textbf{\textcolor{WolfpackRed}{\faClipboardCheck\,Summary of RQ#1.}}}

\title{Learning From the Best: \\ What Makes Popular Hugging Face Models?}
\runningtitle{Learning From the Best: What Makes Popular Hugging Face Models?}

\author{Yinan Wu\affmark{1}, Zhou Yang\affmark{2,3,4}, Bowen Xu\affmark{1},
  Bo Wang\affmark{5}, Bach Le\affmark{6}, David Lo\affmark{7}}
\affiliation{%
  \affmark{1}North Carolina State University, Raleigh, NC, USA\quad
  \affmark{2}University of Alberta, Edmonton, AB, Canada\\
  \affmark{3}Alberta Machine Intelligence Institute, Edmonton, AB, Canada\quad
  \affmark{4}Canada CIFAR AI Chair\\
  \affmark{5}Beijing Jiaotong University, Beijing, China\quad
  \affmark{6}The University of Melbourne, Melbourne, Australia\\
  \affmark{7}Singapore Management University, Singapore}

\emails{\email{ywu92@ncsu.edu},\quad\email{zy25@ualberta.ca},\quad\email{bxu22@ncsu.edu},\quad
  \email{wangbo_cs@bjtu.edu.cn},\quad\email{bach.le@unimelb.edu.au},\quad\email{davidlo@smu.edu.sg}}

\logo{\includegraphics[height=1.0cm]{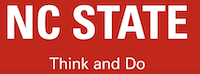}}
\runninglogo{\includegraphics[height=0.65cm]{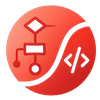}}
\banner{%
  \includegraphics[height=0.85cm]{logo.png}%
  \hspace{0.25cm}%
  \raisebox{0.25cm}{Soft\textcolor{bannerlight}{Max} Lab}}

\codelink{https://github.com/YinanWusoymilk/What-Makes-HF-Models-Popular}{github.com/YinanWusoymilk/What-Makes-HF-Models-Popular}

\begin{document}
\maketitle

\begin{abstract}
Hugging Face has become a central platform for sharing and reusing pre-trained models. As the number of hosted models grows, model practitioners face a practical question: which practices are associated with higher visibility and adoption? Existing studies describe Hugging Face reuse and model-card quality, but they do not systematically compare popular and unpopular models across documentation, repository structure, platform metadata, and technical metadata. In this paper, we studied 66{,}909 Hugging Face models collected on 2026-02-28. We extracted 31 features across four dimensions: Documentation, Models, Platforms, and Techniques. For two popularity indicators, downloads and likes, we first compared the top 10\% popular group with the bottom 80\% unpopular group using a 10\% gap buffer, and assessed feature differences with statistical tests and effect sizes. We then trained five classifiers to predict popularity and used the best-performing model to rank feature importance. We further tested whether the classifier remained reliable across model domains and contributor affiliations. We found that popular models consistently had longer model cards, more code examples, and more GitHub and arXiv links; quantized models showed contrasting popularity patterns under downloads and likes. The best-performing Random Forest reached AUC 0.861 for downloads and 0.848 for likes, with root-level file counts and model size also ranking among the important features for prediction. Cross-group transfer stayed above random in most cases but dropped by 0.12--0.17 AUC on average, showing that prediction performance was partly group-specific. The main findings remained stable after removing the top 1\% download outliers and under three alternative popularity-threshold schemes. A manual review of sampled popular model cards further connected the main documentation features to concrete model-card content. We translate these findings into recommendations for model practitioners and insights for future studies of model-sharing platforms.

\end{abstract}

\section{Introduction}
\label{intro}

Artificial Intelligence (AI) models, especially pre-trained models (PTMs), are now reused across many tasks, including natural language processing (NLP)~\cite{brown2020language,radford2019language}, computer vision (CV)~\cite{he2016deep}, speech recognition~\cite{ardila2019common}, and code generation~\cite{chen2021evaluating}. This reuse is supported by model-sharing platforms, where model practitioners publish model weights, model cards, platform metadata, etc., for others to inspect and reuse. Hugging Face\footnote{\url{https://huggingface.co/}} has become the largest public hub for this practice~\cite{taraghi2024deep,yang2024ecosystem}. On this platform, each shared model is hosted as a model repository. On our data collection date, 2026-02-28, Hugging Face hosted more than 2.6 million public model repositories across many tasks. As the number of models accumulates, visibility becomes a practical concern for practitioners:

\begin{quote}
\textit{How to make models more popular so that they are visible to more users and can have a greater impact?}

\end{quote}
Popularity matters in practice. Hugging Face provides downloads and likes as two popularity indicators. Users can sort models by either indicator when searching the Hub. Models that appear higher in these listings are easier to discover and more likely to be reused as baselines or starting points for downstream tasks~\cite{jiang2023empirical,yang2024ecosystem,jones2024we,shen2023hugginggpt}. On the open-source software (OSS) side, perceived quality and community trust have all been linked to repository popularity indicators such as GitHub stars and forks~\cite{wang2023study,Fan_2021,borges2016understanding}. Practitioners therefore have a clear incentive to understand which practices are associated with higher popularity on Hugging Face, but the evidence available to them is still limited.

Existing research partially fills this gap. Prior OSS studies showed that documentation quality, repository structure, activity, and owner responsiveness are associated with GitHub stars and forks~\cite{borges2016understanding,borges2016predicting,zhu2014patterns,wang2023study}. The closest prior work to ours, by Fan et al.~\cite{Fan_2021}, studied academic AI repositories on GitHub and found that popular repositories differed from unpopular ones in documentation, code, and reproducibility practices.
However, Hugging Face model repositories differ from GitHub-hosted repositories. A Hugging Face model repository is not only a code archive or a README page; it also provides model-specific and platform-specific metadata, such as task tags, implementation libraries, weight files, and Spaces or widgets. These fields shape how models are discovered, loaded, and evaluated on the Hub. Recent Hugging Face studies have characterized model reuse, naming, packaging, and model-card completeness~\cite{jiang2023empirical,jiang2023naming,pepe2024hugging,liang2024s,liang2024systematic,jones2024we}. However, they mainly provide descriptive evidence. They do not identify which individual features are associated with popularity, nor which features are most important for prediction.

In this paper, we addressed this gap with a large-scale empirical study of what makes Hugging Face models popular. We collected public Hugging Face model repositories on 2026-02-28. After removing repositories created within 180 days before data collection, repositories smaller than 300\,MiB, repositories with fewer than 2 likes, and broken or restricted repositories, we retained \textbf{66{,}909 models} for analysis. For each model, we extracted \textbf{31 features} grouped into four dimensions: \textit{Documentation}, \textit{Models}, \textit{Platforms}, and \textit{Techniques}. We measured popularity using two indicators provided by Hugging Face: 30-day \emph{downloads} and cumulative \emph{likes}, and analyzed each indicator independently.

We organize the study around three research questions.

\textbf{RQ1: Which features are associated with model popularity?} We compared the top 10\% (popular) and bottom 80\% (unpopular) models on each indicator, with a 10\% gap buffer, using Mann--Whitney U tests for continuous features and chi-square tests for binary features, with FDR correction. We reported effect sizes using Cliff's $\delta$ for continuous features and Cram\'er's $V$ for binary features. Documentation features accounted for most features with at least small effect: popular models had longer model cards, more code examples, and more GitHub and arXiv links.

\textbf{RQ2: Which features best predicted model popularity?} We filtered redundant features with Spearman correlation, ranked the remaining features by mutual information, trained five classifiers with grid search and repeated stratified cross-validation, and reported permutation importance of the best classifier on a held-out test set. Random Forest performed best, reaching AUC 0.861 for downloads and 0.848 for likes. The most important features were mainly related to documentation depth, repository structure, model size, and quantization.

\textbf{RQ3: Does popularity prediction remain reliable across model domains and contributor affiliations?} We mapped models to task domains using models' metadata and to contributor affiliations using author profile pages. We then evaluated within-group, one-versus-one, and others-versus-one transfer with bootstrap confidence intervals. Within-group AUCs remained above 0.7 across all evaluated domains and contributor affiliations, while cross-group prediction produced lower AUCs, especially across contributor affiliations. Training on all other groups together produced higher AUCs than training on a single source group.

The main contributions of this paper are as follows.

\begin{itemize}
    \item \textbf{Large-scale evidence on Hugging Face popularity.} We studied 66{,}909 Hugging Face models and analyzed downloads and likes separately, showing which documentation, repository, platform, and technical features distinguish the popular group from the unpopular group.
    \item \textbf{Predictive and cross-group analysis.} We trained and interpreted classifiers for popularity prediction under both downloads and likes, and tested whether prediction performance remained reliable across model domains and contributor affiliations. We also evaluated the stability of the results under outlier removal and alternative popularity thresholds.
    \item \textbf{Practical recommendations and replication package.} We translated the findings into practical recommendations for model practitioners and provided a public replication package\footnote{\url{https://github.com/YinanWusoymilk/What-Makes-HF-Models-Popular}} for future studies of model-sharing platforms.
\end{itemize}

The rest of the paper is structured as follows. Section~\ref{sec:background} reviews related work. Section~\ref{sec:data} describes data preparation, including collection, filtering, and labelling. Section~\ref{sec:features} defines the 31 features and the extraction pipeline. Sections~\ref{sec:rq1} to \ref{sec:rq3} report the research questions and the results. Section~\ref{sec:discussion} presents the robustness analyses, the qualitative review of sampled model cards, and practical recommendations. Section~\ref{sec:threats} discusses threats to validity. Section~\ref{sec:conclusion} concludes.

\section{Background and Related Work}\label{sec:background}

\subsection{Popularity in Open-Source Ecosystems}\label{subsec:rw-oss}

Understanding why OSS projects receive more attention than others has been a recurring topic in Mining Software Repositories (MSR) and empirical software engineering. Prior work commonly studied GitHub engagement measures, such as stars, forks, watchers, and contributor counts~\cite{borges2016understanding,borges2016predicting,zhu2014patterns}. These studies connected repository popularity with project features. For example, repositories with conventional folders such as \textit{docs}, \textit{tests}, and \textit{examples} tend to attract more forks~\cite{zhu2014patterns}, while documentation quality, maintenance activity, and owner responsiveness are associated with star growth~\cite{borges2016understanding}. The early growth of GitHub stars can also help predict later popularity~\cite{borges2016predicting}.

Documentation has received particular attention because it is often the first file that potential users inspect. Aggarwal et al.~\cite{aggarwal2014co} found that source code and README content co-evolve in mature projects. Prana et al.~\cite{prana2019categorizing} characterized common README sections, such as purpose, usage, contribution guidelines, and license information, and reported that more complete READMEs are associated with stronger community engagement. Wang et al.~\cite{wang2023study} further studied README content features, including length, code blocks, images, and external links, and related them to GitHub repository popularity. Together, these studies suggest that repository structure and documentation provide useful information about how a project is perceived and adopted.

However, most existing popularity studies were designed for GitHub-hosted code repositories. Hugging Face model repositories provide different types of information from GitHub projects. In addition to model-card content, they include model-specific and platform-specific fields such as task labels, implementation libraries, model-file formats, and demo features.\footnote{\url{https://huggingface.co/docs/hub/model-repos} and \url{https://huggingface.co/docs/hub/api}} These fields may shape how models are discovered and reused, but they have not been studied together with documentation and repository structure in a popularity analysis of model repositories.

\subsection{Sharing PTMs on Hugging Face}\label{subsec:rw-hf}

Hugging Face has become a major platform for sharing PTMs across NLP, CV, audio, and multimodal tasks~\cite{brown2020language,radford2019language,chen2021evaluating,taraghi2024deep,yang2024ecosystem,shen2023hugginggpt}. Recent empirical studies have started to examine this ecosystem. Jiang et al. studied how Hugging Face models are named and versioned~\cite{jiang2023naming}, how users discover and reuse models across downstream tasks~\cite{jiang2023empirical}, and where PTM packaging becomes fragile in practice~\cite{yang2023users}. The PeaTMOSS dataset~\cite{jiang2024peatmoss} further linked PTM repositories with downstream GitHub projects, supporting reproducible studies of model reuse at scale. Wang et al.~\cite{wang2026hidden} studied licensing risks in LLMware supply chains that involve OSS projects, Hugging Face models, and datasets. This line of work shows that Hugging Face model repositories are part of a broader reuse ecosystem, not just storage locations for model weights.

Researchers have also examined the documentation practices around model cards. Mitchell et al.~\cite{mitchell2019model} introduced model cards as a structured way to describe a model's intended use, performance, and limitations. This format later became a common convention on Hugging Face. Pepe et al.~\cite{pepe2024hugging} and Liang et al.~\cite{liang2024s,liang2024systematic} studied how model cards are completed in practice and found large variation in sections such as intended use, evaluation, and limitations. Jones et al.~\cite{jones2024we} proposed a 0 to 5 documentation-quality scale for Hugging Face model cards and found that the top 1,000 most-downloaded models scored higher than the bottom 1,000.

These studies provide the closest Hugging Face context for our work because they characterize model repositories and their documentation and describe reuse, packaging, licensing, and model-card quality. Our study complements this line of work by examining which repository and model-card features are associated with popularity.

\subsection{Feature-Based Approaches to Predicting Project Popularity}\label{subsec:rw-features}

Feature-based analysis has been widely used in MSR to connect project features with project outcomes. Bao et al.~\cite{bao2017will}, for example, used developer activity and working-hour features to predict developer attrition with several classifiers. For popularity, Wang et al.~\cite{wang2023study} extracted README content features from GitHub repositories and found that length, code blocks, and images were informative. The closest prior work to ours is Fan et al.~\cite{Fan_2021}, who studied academic AI repositories on GitHub. They extracted 21 features from documentation, code, and reproducibility dimensions, compared popular and unpopular repositories, and trained random forest classifiers to identify the features that best distinguish the two groups. Their results highlighted actionable practices such as providing richer README content, linking related repositories, including images, and adding a license.

Our study builds on this feature-based research paradigm, but shifts the setting from GitHub-hosted repositories to Hugging Face model repositories. Instead of GitHub stars, we used downloads and likes as popularity indicators. We extracted \textbf{31 features} across four dimensions: \textit{Documentation}, \textit{Models}, \textit{Platforms}, and \textit{Techniques}. Several of these features are specific to model-sharing platforms, including platform metadata, task and library information, model-file formats, and file-tree structures. We also extended the analysis beyond a single classification task by testing feature-level differences (RQ1), ranking predictive features with model-based importance (RQ2), and evaluating cross-domain and cross-affiliation performance (RQ3). To the best of our knowledge, this is the first feature-based study that connected Hugging Face-specific features to model popularity and examined whether popularity prediction remains reliable across model domains and contributor affiliations.

\section{Data Preparation}\label{sec:data}

In this section, we describe how we collected the Hugging Face model repositories, filtered them for analysis, and defined the downloads- and likes-based popularity groups.

\subsection{Data Collection}\label{subsec:snapshot}

We collected the data on 2026-02-28. The initial crawl returned 2{,}669{,}509 public model repositories from the Hugging Face Hub. We collected the data in three steps.

First, we used the \texttt{/api/models} endpoint to enumerate public model repositories. The crawler started from \path|/api/models?limit=1000|, followed the pagination URL in \texttt{response.links["next"]}, and stored the metadata returned for each model, including basic fields such as the model identifier, creation time, downloads, and likes. Second, we queried two repository-level endpoints for each model: \path|/api/models/{model_id}| for per-model details, and \path|/api/models/{model_id}/tree/main| for the repository file tree and per-file byte sizes. Third, we downloaded the raw \texttt{README.md} from each model repository using the \path|/{model_id}/raw/main/README.md| path.

\subsection{Filtering}\label{subsec:filtering}

After removing 9 duplicate \texttt{model\_id} entries, we obtained 2{,}669{,}500 unique model repositories. We then applied four filters to construct the final dataset. First, we removed repositories created within 180 days before data collection, i.e., on or after 2025-09-01. Recently created models may not have had enough time to accumulate stable downloads or likes, which would have introduced time-on-platform noise into the popularity labels. Second, we observed that repositories without model files were
often small in size, thus we filtered out those repositories
that are smaller than 300\,MiB. Third, we removed repositories with fewer than 2 likes. Repositories with 0 or 1 likes provided little ranking information, and the heavy tie at the bottom of the distribution would have made the unpopular group less meaningful. Finally, we removed repositories whose Hub API responses indicated a broken or gated/private repository (\texttt{if\_repository}~$\ne 1$ or \texttt{if\_restricted}~$\ne 0$). This final filter did not remove additional models in our collected data. After filtering, the final dataset contained \textbf{66{,}909 models}. Table~\ref{tab:filter-waterfall} shows the number of models remaining after each step.

\begin{table}[t]
\centering
\small
\caption{Filtering steps from the 2026-02-28 Hugging Face crawl to the final dataset. Each row reports the number of models remaining after the listed filter is applied; filters are applied cumulatively in the order shown.}
\label{tab:filter-waterfall}
\begin{tabular}{l r}
\hline
\textbf{Step} & \textbf{Models remaining} \\
\hline
Raw crawl (2026-02-28, public repositories) & 2{,}669{,}509 \\
After deduplicating \texttt{model\_id} & 2{,}669{,}500 \\
\texttt{created\_time} $<$ 2025-09-01 (180-day recency) & 1{,}935{,}844 \\
\texttt{model-size} $\geq$ 300\,MiB & 967{,}002 \\
\texttt{likes} $\geq$ 2 & 66{,}909 \\
Valid, non-restricted repository & 66{,}909 \\
\hline
\textbf{Final dataset} & \textbf{66{,}909} \\
\hline
\end{tabular}
\end{table}

\subsection{Data Labelling}\label{subsec:labelling}

We labelled each model repository as popular or unpopular using two popularity indicators reported by the Hugging Face Hub: the cumulative number of \textbf{likes} and the monthly number of \textbf{downloads}. Likes reflect explicit user appreciation, while downloads also capture automated retrievals from downstream code paths. 
The two indicators were related but not interchangeable. On the final 66{,}909-model dataset, their Spearman rank correlation was $\rho = 0.36$. Under the interpretation used by Wang et al.~\cite{wang2023study} ($|\rho| < 0.3$ weak; $0.3 \leq |\rho| < 0.7$ moderate; $|\rho| \geq 0.7$ strong), this indicates a moderate association. We therefore treated downloads and likes as two separate popularity indicators in the labeling step.

\begin{figure}[htbp]
\centering
\begin{minipage}[t]{0.485\textwidth}
  \centering
  \includegraphics[width=\linewidth]{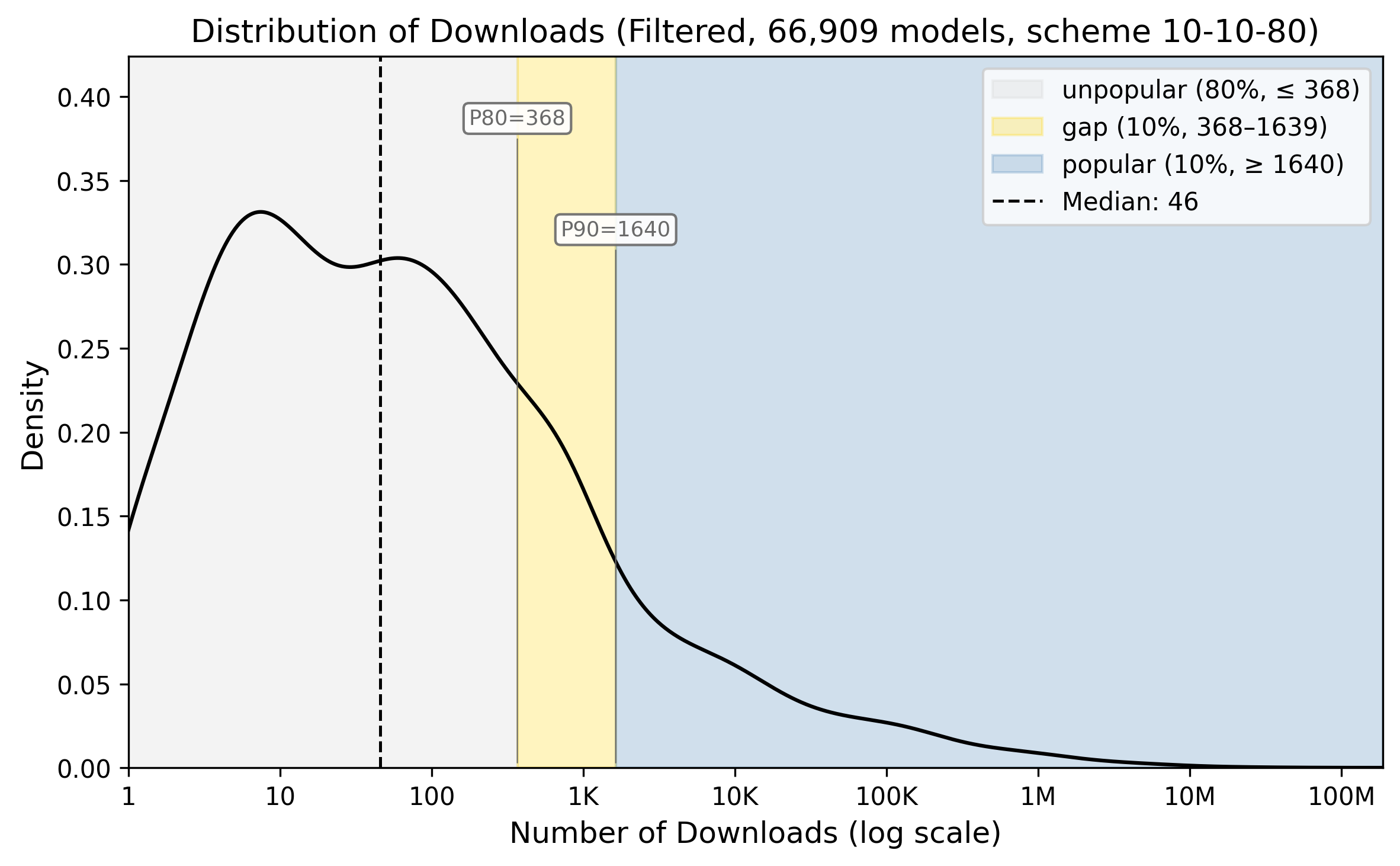}\\
  {\small (a) Downloads\par}
\end{minipage}\hfill
\begin{minipage}[t]{0.485\textwidth}
  \centering
  \includegraphics[width=\linewidth]{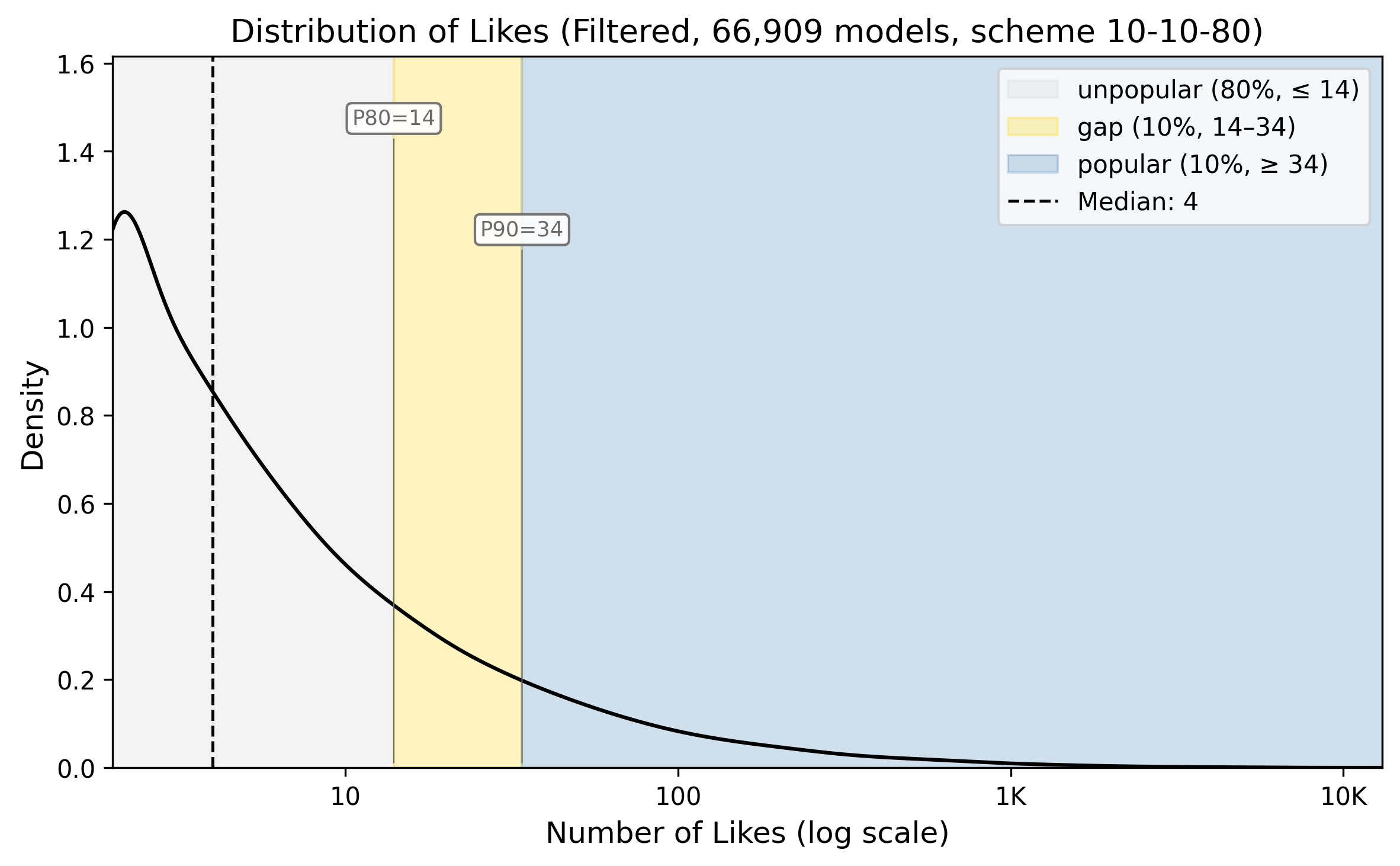}\\
  {\small (b) Likes\par}
\end{minipage}
\caption{Distribution of the two popularity indicators on the 66{,}909-model final dataset under the 10-10-80 scheme. The horizontal axis is log-scaled.}
\label{fig:distribution}
\end{figure}

For each indicator, we sorted the 66{,}909 models in descending order and applied a \textbf{10-10-80} scheme: the top 10\% (6{,}690 models) were labeled \textit{popular}, the bottom 80\% (53{,}528 models) were labeled \textit{unpopular}, and the 10\% in between (6{,}691 models) was used as a gap buffer. The gap buffer made the difference between the popular and unpopular groups more substantial, following the same rationale used in prior study~\cite{Fan_2021}. Figure~\ref{fig:distribution} shows the resulting partition for downloads and likes.

\section{Feature Extraction}\label{sec:features}

We extracted \textbf{31 features} for each model repository and grouped them into four dimensions: \textit{Documentation}, \textit{Models}, \textit{Platforms}, and \textit{Techniques}. Table~\ref{tab:features} summarizes the feature definitions, and the following subsections describe how the features in each dimension were extracted.

\begin{table}[!htbp]
  \centering
  \caption{The 31 features analyzed in our study, organised into four dimensions.}
  \scriptsize
  \begin{tabularx}{\textwidth}{p{2.2cm}p{3.0cm}X}
      \hline
      \textbf{Dimension} & \textbf{Feature} & \textbf{Description} \\ \hline
      \multirow{14}{*}{Documentation}
       & \texttt{length-yaml} & The total number of words in the YAML part of the README. \\
       & \texttt{length-doc} & The total number of words in the Markdown part of the README. \\
       & \texttt{has-video} & Whether the documentation includes URLs to videos. \\
       & \texttt{num-inline-code} & The number of inline code snippets in the documentation. \\
       & \texttt{num-code-block} & The number of fenced code blocks in the documentation. \\
       & \texttt{num-static-img} & The number of static images (e.g., \texttt{.jpg}, \texttt{.png}). \\
       & \texttt{num-animation} & The number of animated images (e.g., \texttt{.gif}). \\
       & \texttt{num-lists} & The number of lists used in the documentation. \\
       & \texttt{num-tables} & The number of tables included in the documentation. \\
       & \texttt{num-gh-links} & The number of GitHub links in the documentation. \\
       & \texttt{num-hf-links} & The number of Hugging Face links in the documentation. \\
       & \texttt{num-arxiv-links} & The number of arXiv identifiers declared in the metadata. \\
       & \texttt{has-bibtex} & Whether the README contains a BibTeX entry. \\
       & \texttt{has-license} & Whether license information is declared in the metadata. \\
      \hline
      \multirow{10}{*}{Models}
       & \texttt{has-config} & Whether the model repository contains a configuration file. \\
       & \texttt{has-results} & Whether the metadata reports performance results (\texttt{model-index}). \\
       & \texttt{has-dataset} & Whether the metadata specifies any dataset information. \\
       & \texttt{num-dataset} & The number of datasets linked to the model. \\
       & \texttt{match-hf-dataset} & Whether the linked datasets are hosted on Hugging Face. \\
       & \texttt{model-size} & The total size of the model files in the repository. \\
       & \texttt{num-root-files} & The number of files in the root directory. \\
       & \texttt{num-modules} & The number of modules (subdirectories) in the repository. \\
       & \texttt{num-model-files} & The number of model-weight files (\texttt{.bin}, \texttt{.safetensors}, etc.). \\
       & \texttt{has-quantized} & Whether the model is a quantized variant. \\
      \hline
      \multirow{3}{*}{Platforms}
       & \texttt{has-space} & Whether the model is featured in at least one Hugging Face Space. \\
       & \texttt{has-safetensors} & Whether the model uses SafeTensors to store model weights. \\
       & \texttt{has-widgets} & Whether the model's inference API contains widget data. \\
      \hline
      \multirow{4}{*}{Techniques}
       & \texttt{has-pipeline} & Whether the metadata specifies a named pipeline (\texttt{pipeline\_tag}). \\
       & \texttt{has-impl-lib} & Whether the metadata specifies a primary implementation library. \\
       & \texttt{has-supported-lib} & Whether the model uses a library officially supported by Hugging Face. \\
       & \texttt{has-additional-fw} & Whether the model supports additional frameworks beyond the primary one. \\
      \hline
  \end{tabularx}
  \label{tab:features}
\end{table}

\subsection{Documentation Dimension}\label{subsec:doc-dim}

Documentation is the main way in which model practitioners explain what a model does, how it should be used, and how it should be cited. Prior studies~\cite{wang2023study,Fan_2021,prana2019categorizing,aggarwal2014co} have similarly treated README content as an important indicator of project quality and adoption. On Hugging Face, each model card is stored in the repository as \texttt{README.md} and rendered on the model page. It usually contains a Markdown body and, when available, a YAML metadata header.

Following Fan et al.~\cite{Fan_2021}, we parsed the README into HTML before extracting documentation features. We used Mistune\footnote{\url{https://mistune.lepture.com/}} as the main Markdown parser and Pandoc\footnote{\url{https://pandoc.org/}} for table detection.
We first separated the YAML metadata block from the Markdown body and counted words in both parts. These counts gave \texttt{length-yaml} and \texttt{length-doc}. We then counted structural elements in the rendered body: \texttt{num-lists} was the number of \texttt{<ol>} and \texttt{<ul>} elements, and \texttt{num-tables} was the number of \texttt{<table>} elements in the Pandoc-rendered HTML.

Runnable examples are especially important for model repositories, where unclear usage instructions are a common source of user issues~\cite{yang2023users}. We counted code by inspecting each \texttt{<code>} element. If its parent was \texttt{<pre>}, we counted it as a fenced code block (\texttt{num-code-block}); otherwise, we counted it as inline code (\texttt{num-inline-code}). 
We also counted images and videos. Static images were counted when an \texttt{<img>} source used an image extension such as \texttt{.png}, \texttt{.jpg}, or \texttt{.svg}; animated images were counted when the source ended in \texttt{.gif}. The binary feature \texttt{has-video} was set to 1 when a link or iframe pointed to a video source such as YouTube or a video file.

Finally, we captured links and citation-related metadata, because links to code, papers, and citation entries help users trace a model's provenance and additional materials. We counted GitHub and Hugging Face links by inspecting the \texttt{href} attributes of \texttt{<a>} and \texttt{<link>} elements. We counted arXiv references from the metadata returned by \texttt{/api/models} rather than from README text, because Hugging Face returns them as structured \texttt{arxiv:<paper-id>} tags. The feature \texttt{num-arxiv-links} was therefore the number of such arXiv tag entries. We set \texttt{has-bibtex} to 1 when the README contained a BibTeX code block, either through a \texttt{language-bibtex} class or through a code block containing the standard fields \texttt{author=}, \texttt{title=}, and \texttt{year=}. We set \texttt{has-license} to 1 when the metadata included a \texttt{license:<spdx-identifier>} tag.

\subsection{Models Dimension}\label{subsec:mod-dim}

The Models dimension captures properties of the model files and repository layout. These features are specific to model-sharing platforms and are mostly absent from prior GitHub-based popularity studies~\cite{jones2024we,Fan_2021}. We extracted them from two sources: metadata, the per-model details and repository file-tree endpoints.

We used the per-model details endpoint to describe repository structure, because the repository structure may affect how easily practitioners can inspect, load, and reuse a model.
The endpoint returned a \texttt{siblings} list that enumerated repository files. We counted the files at the repository root and used this count as \texttt{num-root-files}. For files in subdirectories, we collected the first path component and reported the number of unique components as \texttt{num-modules}. We also counted model-weight files with common weight-file extensions, such as \texttt{.bin}, \texttt{.safetensors}, and \texttt{.onnx}. The feature \texttt{has-config} was set to 1 when the per-model details JSON contained a non-empty \texttt{config} field. We computed \texttt{model-size} by summing the per-file byte sizes returned by the repository tree endpoint, which avoided downloading the model weights.

We captured dataset and result reporting because these fields may help practitioners identify whether a model is relevant and trustworthy. The feature \texttt{has-dataset} was set to 1 if the structured metadata contained at least one \texttt{dataset:<name>} entry, and \texttt{num-dataset} counted those entries. We set \texttt{match-hf-dataset} to 1 if at least one declared dataset appeared in the set of Hugging Face dataset identifiers collected in Section~\ref{sec:data}. We set \texttt{has-results} to 1 when the per-model details contained a non-empty \texttt{model-index} block with at least one entry containing a non-empty \texttt{results} field.

Quantized models are often smaller or faster than full-precision checkpoints~\cite{han2015deep}, which may make them easier to deploy and share. We therefore extracted whether a model was quantized. Hugging Face did not expose a single canonical quantization flag, so we combined model identifiers and tag information. We searched for common quantization terms from the Hugging Face quantization documentation,\footnote{\url{https://huggingface.co/docs/transformers/main/en/quantization/selecting}} such as \texttt{gguf}, \texttt{gptq}, and \texttt{4bit}. We also accepted the exact tag \texttt{quantized} and the substring \texttt{quantized} in \texttt{model\_id}, while excluding negated forms such as \texttt{unquantized} and \texttt{non-quantized}. A model was assigned \texttt{has-quantized = 1} when any of these checks matched.

\subsection{Platforms Dimension}\label{subsec:plat-dim}

The Platforms dimension captures features that make a model easier to inspect or use from the Hugging Face interface. We included three binary features. The feature \texttt{has-safetensors} was set to 1 when the structured metadata contained a \texttt{safetensors}-prefixed entry. SafeTensors is a model-weight format designed to avoid the code-loading risks associated with pickle-based weight files. The feature \texttt{has-space} was set to 1 when the metadata indicated that the model was connected to a Hugging Face Space. Spaces let users try a model in the browser and may improve discoverability~\cite{shen2023hugginggpt}. The feature \texttt{has-widgets} was set to 1 when the model details contained a non-empty \texttt{widgetData} field, which enables example inputs on the model card.

\subsection{Techniques Dimension}\label{subsec:tech-dim}

The Techniques dimension describes how a model is implemented and made available to users. All four features were extracted from structured metadata. We set \texttt{has-pipeline} to 1 when the model entry contained a non-empty \texttt{pipeline\_tag}. A pipeline tag links the model to a Hugging Face task abstraction, which helps users find the model by task and load it through standard interfaces. We set \texttt{has-impl-lib} to 1 when the entry contained a non-empty \texttt{library\_name}, such as \texttt{transformers}, \texttt{diffusers}, or \texttt{sentence-transformers}. We set \texttt{has-additional-fw} to 1 when the metadata listed an additional framework or library beyond the primary \texttt{library\_name}, such as \texttt{pytorch}, \texttt{tensorflow}, or \texttt{onnx}.

\subsection{Manual Sanity Check}\label{subsec:sanity-check}

To check the correctness of the extraction pipeline, we manually verified a stratified random sample of 30 models. One author compared and verified the Hugging Face model page with the extracted values for all 31 features. We excluded \texttt{downloads} and \texttt{likes} from this check because they are live Hub counters and can change after data collection. 

\section{RQ1: Features Associated with Model Popularity}\label{sec:rq1}

\subsection{Motivation}\label{subsec:rq1-motivation}

Following prior studies~\cite{Fan_2021,wang2023study}, we first examined how individual features differed between popular and unpopular models. RQ1 compared the 31 extracted features under downloads and likes separately, allowing us to see whether the two popularity indicators pointed to the same feature dimensions.

\subsection{Method}\label{subsec:rq1-method}

We compared the \textit{popular} and \textit{unpopular} groups defined under the main 10-10-80 scheme (Section~\ref{subsec:labelling}), excluding the gap buffer. For each of the 16 continuous features, we used the Mann--Whitney U test~\cite{mann1947test} and reported the signed Cliff's $\delta$; positive values indicate that the feature tended to be larger in the popular group. For each of the 15 binary features, we used the Chi-square test and reported Cramér's $V$. Since Cramér's $V$ is non-directional, we indicated direction by comparing feature prevalence between the two groups. We applied the Benjamini--Hochberg false discovery rate procedure~\cite{benjamini1995controlling} at $\alpha = 0.05$ to the 31 tests under each indicator. We interpreted effect-size magnitudes using standard thresholds: for Cliff's $\delta$, negligible ($<0.147$), small ($0.147$ to $0.33$), medium ($0.33$ to $0.474$), and large ($\geq 0.474$)~\cite{romano2006appropriate}; for Cramér's $V$, negligible ($<0.1$), small ($0.1$ to $0.3$), medium ($0.3$ to $0.5$), and large ($\geq 0.5$)~\cite{cohen2013statistical}. The full analysis was run twice, once using downloads-based labels and once using likes-based labels.

\subsection{Results}\label{subsec:rq1-result}

After FDR correction at $\alpha = 0.05$, \textbf{28 of 31 features} were statistically significant under the downloads indicator, and \textbf{29 of 31 features} were statistically significant under the likes indicator. However, most statistically significant differences were negligible in magnitude. Under the thresholds above, \textbf{13 of 31 features reached at least small effect on downloads}, and \textbf{8 of 31 features reached at least small effect on likes}. Table~\ref{tab:rq1-small-effects} reports the union of the 15 features that reached at least small effect on at least one indicator.

\begin{table}[!htbp]
\centering
\small
\caption{Features that reach at least small effect on at least one popularity indicator. All rows shown are FDR-significant. Continuous features report signed Cliff's $\delta$; positive values mean larger values in the popular group. Binary features report Cramér's $V$, with $\uparrow$ / $\downarrow$ indicating whether the feature is more / less prevalent in the popular group. \textbf{Bold} = small or larger.}
\label{tab:rq1-small-effects}
\begin{tabular}{l l r r}
\hline
\textbf{Dimension} & \textbf{Feature} & \textbf{downloads} & \textbf{likes} \\
\hline
Documentation & \texttt{length-yaml} & \textbf{+0.152} & +0.049 (negligible) \\
Documentation & \texttt{length-doc} & \textbf{+0.296} & \textbf{+0.292} \\
Documentation & \texttt{num-code-block} & \textbf{+0.282} & \textbf{+0.252} \\
Documentation & \texttt{num-inline-code} & \textbf{+0.170} & \textbf{+0.180} \\
Documentation & \texttt{num-lists} & +0.124 (negligible) & \textbf{+0.190} \\
Documentation & \texttt{num-tables} & \textbf{+0.180} & +0.117 (negligible) \\
Documentation & \texttt{num-gh-links} & \textbf{+0.264} & \textbf{+0.227} \\
Documentation & \texttt{num-arxiv-links} & \textbf{+0.268} & \textbf{+0.249} \\
Documentation & \texttt{has-bibtex} & \textbf{0.183~$\uparrow$} & \textbf{0.152~$\uparrow$} \\
Models & \texttt{has-config} & \textbf{0.116~$\uparrow$} & 0.065~$\uparrow$ (negligible) \\
Models & \texttt{has-quantized} & 0.032~$\downarrow$ (negligible) & \textbf{0.113~$\downarrow$} \\
Platforms & \texttt{has-safetensors} & \textbf{0.107~$\uparrow$} & 0.069~$\uparrow$ (negligible) \\
Techniques & \texttt{has-pipeline} & \textbf{0.128~$\uparrow$} & 0.098~$\uparrow$ (negligible) \\
Techniques & \texttt{has-impl-lib} & \textbf{0.107~$\uparrow$} & 0.049~$\uparrow$ (negligible) \\
Techniques & \texttt{has-additional-fw} & \textbf{0.121~$\uparrow$} & 0.030~$\uparrow$ (negligible) \\
\hline
\end{tabular}
\end{table}

\textbf{Documentation accounted for most small-effect differences.} Documentation contributed the largest share of small-effect features on both indicators: 8 of the 14 Documentation features reached small effect on downloads and 7 of 14 on likes. Six documentation features did so under both indicators: \texttt{length-doc}, \texttt{num-code-block}, \texttt{num-inline-code}, \texttt{num-gh-links}, \texttt{num-arxiv-links}, and \texttt{has-bibtex}. Popular model cards were longer, contained more code blocks and inline code, included more GitHub and arXiv links, and more often included a BibTeX block. The largest single difference was \texttt{length-doc}: popular cards had a median of 451 words versus 234 under downloads, and 483 versus 240 under likes. These differences are consistent with how practitioners inspect model cards: longer cards, runnable examples, links to code or papers, and citation blocks help users understand what a model does, try it, trace its source, and cite it. A few documentation features were indicator-specific: \texttt{length-yaml} and \texttt{num-tables} reached small effect only under downloads, while \texttt{num-lists} reached small effect only under likes.

\textbf{Technique features reached small effects mainly under downloads.} Under downloads, three of the four Technique features reached small effect: \texttt{has-pipeline}, \texttt{has-impl-lib}, and \texttt{has-additional-fw}. Under likes, all four Technique features remained negligible. This result is consistent with the role of technical metadata on the Hub: pipeline and library declarations can make a model easier for downstream applications and scripts to find and load, while they are less directly tied to explicit user appreciation.

\textbf{Model and platform features showed fewer small-effect differences.} In the Models dimension, only \texttt{has-config} reached small effect under downloads. This is consistent with configuration files making a repository easier for Hugging Face tooling and downstream scripts to load. Most other model-structure features remained negligible. The notable exception was \texttt{has-quantized}, which was more prevalent in the unpopular group under likes: 14.7\% of popular models were quantized, compared with 31.1\% of unpopular models ($V = 0.113$). This does not imply that quantization reduces popularity; rather, it suggests that quantized releases may follow a different popularity pattern, especially under likes. In the Platforms dimension, \texttt{has-safetensors} reached small effect under downloads but remained negligible under likes, which is consistent with SafeTensors being primarily a loading and model-file format feature.

\begin{rqsummarybox}
\rqsummarylabel{1} Popular and unpopular models differed on many features after FDR correction, but the non-negligible differences were concentrated in a smaller set: 13 features under downloads and 8 under likes reached at least small effect. Documentation was the most consistent dimension across both indicators. Technique metadata appeared mainly in the downloads comparison, while quantization was the only small-effect feature that was more prevalent in the unpopular group, under likes.
\end{rqsummarybox}

\section{RQ2: Feature Importance for Popularity Prediction}\label{sec:rq2}

\subsection{Motivation}\label{subsec:rq2-motivation}

RQ1 compared popular and unpopular models one feature at a time. RQ2 examined the features jointly: we trained classifiers to distinguish popular from unpopular models and used permutation importance to identify which features contributed most to the best classifier.

\subsection{Method}\label{subsec:rq2-method}

We ran the same four-stage procedure separately for downloads and likes, using the popular and unpopular groups defined by the 10-10-80 scheme.

\textit{Stage~1 -- Correlation filtering.} We computed pairwise Spearman correlation $\rho$ across the 31 features and, for every pair with $|\rho| > 0.7$~\cite{guyon2003introduction,vergara2014review}, removed the feature with the lower absolute correlation to the popular/unpopular outcome. 
This step removed five features under each indicator: \texttt{has-dataset}, \texttt{num-dataset}, \texttt{num-model-files}, and \texttt{has-supported-lib}. It also removed one member of the framework/library pair: \path|has-impl-lib| on downloads and \texttt{has-additional-fw} on likes. The remaining 26 features were used in the next stage.

\textit{Stage~2 -- Mutual-information ranking and $k$-selection.} Using the 26 retained features, we computed mutual information~\cite{kraskov2004estimating} with the binary popular/unpopular outcome and ranked features in descending order. We then swept $k$ from 5 to 26, trained a Random Forest~\cite{breiman2001random} on the top-$k$ features for each $k$, and selected the $k$ that maximised mean 5-fold cross-validation ROC AUC. The optimal $k$ was 26 under both indicators, with 5-fold CV AUC of 0.854 on downloads and 0.841 on likes.

\textit{Stage~3 -- Classifier comparison.} We compared five classifiers: Random Forest, Decision Tree, Linear SVM, Gaussian Naive Bayes, and $k$-Nearest Neighbours. Hyperparameters were tuned with \texttt{GridSearchCV}. Each model was evaluated with 10-fold cross-validation repeated 10 times (100 folds total). ROC AUC was the primary metric because it evaluates ranking quality and is less sensitive to the decision threshold under class imbalance~\cite{borges2016predicting,bao2017will}. Accuracy, precision, recall, and F1 were reported as secondary metrics. Following prior popularity-prediction work, we treated AUC $> 0.7$ as the acceptability threshold~\cite{Fan_2021}. Since the popular and unpopular groups had an approximately 1:8 class ratio, we used balanced class weights for classifiers that supported them.

\textit{Stage~4 -- Permutation importance.} For the best classifier, we computed permutation importance on a stratified 80/20 holdout test set with 30 shuffles per feature and ROC AUC as the scoring metric. The reported $\Delta$AUC is the mean drop in test AUC when a feature's values are shuffled. This complements RQ1 because it estimates the importance of each feature conditional on the other features in the model.

\subsection{Results}\label{subsec:rq2-result}

\textbf{Random Forest performed best under both popularity indicators.} Table~\ref{tab:rq2-classifiers} reports the 100-fold scores. Random Forest achieved the highest AUC on both downloads ($0.861 \pm 0.007$) and likes ($0.848 \pm 0.006$). KNN ranked second, Decision Tree third, Linear SVM fourth, and Gaussian Naive Bayes had the lowest AUC under both indicators. All classifiers except Gaussian Naive Bayes exceeded 0.7 AUC, a conventional lower bound for acceptable discrimination used in prior popularity-prediction work~\cite{Fan_2021}. This result is consistent with the use of Random Forest in prior study~\cite{Fan_2021}: the model can combine numeric and binary features, capture non-linear feature combinations, and reduce the instability of a single decision tree through ensembling~\cite{breiman2001random}.

The class imbalance made accuracy alone misleading. An always-unpopular classifier would obtain AUC $= 0.5$ and F1 $= 0$, but still reach 0.889 accuracy because unpopular models make up most of the evaluation set. Random Forest improved only modestly over this baseline in accuracy, but it achieved much higher AUC and non-zero recall for the popular class. On downloads, Random Forest reached precision $= 0.608$, recall $= 0.437$, and F1 $= 0.508$; on likes, it showed higher precision and lower recall (precision $= 0.676$, recall $= 0.232$, F1 $= 0.345$). Decision Tree and Linear SVM showed the opposite pattern, with higher recall but lower precision.

\begin{table}[!htbp]
\centering
\small
\caption{Classifier performance (mean $\pm$ std across 100 folds = 10-fold $\times$ 10 repeats). Best model under each indicator is in bold. ``Always unpopular'' is a majority-class baseline that predicts every model as unpopular; its precision, recall, and F1 for the popular class are reported as 0 because no popular instances are predicted.}
\label{tab:rq2-classifiers}
\begin{tabular}{l c c c c c}
\hline
\textbf{Classifier} & \textbf{AUC} & \textbf{Acc.} & \textbf{Prec.} & \textbf{Rec.} & \textbf{F1} \\
\hline
\multicolumn{6}{l}{\textit{by downloads}} \\
\textit{Always unpopular} & 0.500 & 0.889 & 0.000 & 0.000 & 0.000 \\
\textbf{Random Forest}  & \textbf{0.861} & 0.906 & 0.608 & 0.437 & 0.509 \\
$k$-NN                  & 0.807 & 0.905 & 0.618 & 0.384 & 0.473 \\
Decision Tree           & 0.780 & 0.738 & 0.256 & 0.714 & 0.377 \\
Linear SVM              & 0.722 & 0.659 & 0.194 & 0.658 & 0.300 \\
Gaussian Naive Bayes    & 0.684 & 0.152 & 0.114 & 0.981 & 0.205 \\
\hline
\multicolumn{6}{l}{\textit{by likes}} \\
\textit{Always unpopular} & 0.500 & 0.889 & 0.000 & 0.000 & 0.000 \\
\textbf{Random Forest}  & \textbf{0.848} & 0.902 & 0.676 & 0.232 & 0.345 \\
$k$-NN                  & 0.782 & 0.898 & 0.590 & 0.258 & 0.359 \\
Decision Tree           & 0.758 & 0.712 & 0.234 & 0.700 & 0.350 \\
Linear SVM              & 0.724 & 0.640 & 0.190 & 0.685 & 0.297 \\
Gaussian Naive Bayes    & 0.684 & 0.395 & 0.141 & 0.864 & 0.242 \\
\hline
\end{tabular}
\end{table}

\textbf{Permutation importance identified eight features that were stable across indicators.} Table~\ref{tab:rq2-permimp} lists the top-10 permutation importances on the best Random Forest under each indicator. Eight features appeared in both top-10s: \texttt{length-doc}, \texttt{length-yaml}, \texttt{num-code-block}, \texttt{num-gh-links}, \texttt{num-arxiv-links}, \texttt{model-size}, \texttt{num-root-files}, and \texttt{has-quantized}. These eight common features consisted of five Documentation features and three Models features. \texttt{length-doc} was the strongest feature under both indicators, with $\Delta\text{AUC} = 0.035$ on downloads and $0.067$ on likes. This result reinforces the RQ1 pattern that model-card content remains central even after metadata, file structure, and model-format features are considered together. At the same time, RQ2 showed a result that RQ1 alone would miss: \texttt{model-size} had a negligible feature-level effect in RQ1, but ranked 2nd on downloads and 3rd on likes in permutation importance. This suggests that model size becomes informative when interpreted jointly with documentation and repository-structure features, rather than as a standalone difference between the two groups. Because these rankings come from the best-performing Random Forest, they should be read as model-specific importance estimates rather than model-independent causal effects.

The stronger predictive performance in RQ2 therefore does not contradict the small effect sizes in RQ1. RQ1 measured each feature separately, so a small effect means that the popular and unpopular groups still overlap substantially on that single feature. RQ2 instead evaluated combinations of features after removing highly correlated pairs. The 26 features covered different aspects of a model repository, including model-card detail, external references, file-tree structure, model size, and model format. Small differences across these aspects can accumulate, and Random Forest can capture threshold-based and non-linear combinations that are not visible in one-feature-at-a-time comparisons.

\begin{table}[!htbp]
\centering
\small
\caption{Top-10 permutation importance ($\Delta$AUC, mean $\pm$ std over 30 shuffles) on the best Random Forest under each popularity indicator. \textbf{Bold} = appears in the top-10 on both indicators.}
\label{tab:rq2-permimp}
\begin{tabular}{r l r l r}
\hline
\textbf{Rank} & \textbf{Feature (downloads)} & \textbf{$\Delta$AUC} & \textbf{Feature (likes)} & \textbf{$\Delta$AUC} \\
\hline
1  & \textbf{\texttt{length-doc}}          & 0.035 & \textbf{\texttt{length-doc}}          & 0.067 \\
2  & \textbf{\texttt{model-size}}          & 0.027 & \textbf{\texttt{has-quantized}}       & 0.058 \\
3  & \textbf{\texttt{num-gh-links}}        & 0.024 & \textbf{\texttt{model-size}}          & 0.031 \\
4  & \texttt{has-additional-fw}            & 0.024 & \textbf{\texttt{num-arxiv-links}}     & 0.025 \\
5  & \textbf{\texttt{length-yaml}}         & 0.022 & \textbf{\texttt{length-yaml}}         & 0.020 \\
6  & \textbf{\texttt{num-arxiv-links}}     & 0.019 & \texttt{num-hf-links}                 & 0.019 \\
7  & \textbf{\texttt{num-code-block}}      & 0.013 & \textbf{\texttt{num-code-block}}      & 0.016 \\
8  & \textbf{\texttt{num-root-files}}      & 0.013 & \textbf{\texttt{num-gh-links}}        & 0.016 \\
9  & \textbf{\texttt{has-quantized}}       & 0.011 & \texttt{has-license}                  & 0.015 \\
10 & \texttt{has-config}                   & 0.011 & \textbf{\texttt{num-root-files}}      & 0.012 \\
\hline
\end{tabular}
\end{table}

\textbf{The two indicators shared a common core but differed on secondary features.} The largest cross-indicator difference was \texttt{has-quantized}, which ranked 9th on downloads ($\Delta\text{AUC} = 0.011$) but 2nd on likes ($\Delta\text{AUC} = 0.058$). This result is consistent with RQ1: quantized models were more common in the unpopular group under likes, and the Random Forest used this feature to separate unpopular models from popular ones. One interpretation is that quantized releases often function as deployment-oriented versions of already-known base models. They can receive downloads through reuse, but they do not necessarily receive likes as independent model cards. Two other indicator-specific features also reflect the different meanings of downloads and likes: \texttt{has-additional-fw} ranked 4th on downloads and was removed in Stage~1 for the likes indicator in favour of \texttt{has-impl-lib}, while \texttt{has-license} appeared only in the likes top-10. In contrast, the common top features were mostly about documentation depth, repository structure, and model size, suggesting that both indicators were associated with models that are easier to inspect and reuse.

\begin{rqsummarybox}
\rqsummarylabel{2} A Random Forest trained on the 26 features retained after Spearman filtering and MI selection reached AUC $= 0.861$ on downloads and $0.848$ on likes, outperforming both the 0.7 reference threshold and the always-unpopular baseline. The same eight features appeared in the permutation-importance top-10 under both indicators: five Documentation features and three Models features. \texttt{length-doc} was the strongest feature under both indicators, while \texttt{has-quantized} was much more important under likes than under downloads. Compared with RQ1, RQ2 also showed that some features with negligible feature-level effects, especially \texttt{model-size}, became important when considered jointly with the other features.
\end{rqsummarybox}

\section{RQ3: Cross-Group Generalization of Popularity Prediction}\label{sec:rq3}

\subsection{Motivation}\label{subsec:rq3-motivation}

RQ2 evaluated prediction with all models in the main analysis pooled together, but model repositories may follow different conventions across domains and contributor affiliations. RQ3 therefore examined whether the best RQ2 classifier remained reliable across these groups. This analysis helps assess whether prediction performance is broadly transferable, or whether it depends on domain- or affiliation-specific release practices.

\subsection{Method}\label{subsec:rq3-method}

We added two grouping variables to each model and reused the RQ2 Random Forest with the same 26 features and tuned parameters. Keeping the model setup fixed allowed us to focus on how group differences affected prediction performance.

\textit{Grouping by domain.} We mapped the \texttt{pipeline\_tag} field to six task families aligned with the Hugging Face task taxonomy:\footnote{\url{https://huggingface.co/tasks}} NLP, CV, Multimodal, Audio, Tabular, and Reinforcement Learning. Models with a missing or unmapped \texttt{pipeline\_tag} were assigned to \texttt{Other}. In the current data, \texttt{Other} mostly consists of models without a \texttt{pipeline\_tag}; we therefore interpret it as a mixed missing-tag group rather than as a single task domain.

\textit{Grouping by affiliation.} We parsed each unique contributor's Hugging Face profile page for the \texttt{<span class=``capitalize''>} badge. We retained the badge categories used by the platform: \texttt{company}, \texttt{university}, \texttt{non-profit}, \texttt{community}, \texttt{classroom}, and \texttt{government}. Profiles that loaded but did not display one of these badges were assigned to \texttt{organization or individual}, the platform's default category for accounts without an organization badge. Profiles that could not be fetched were assigned to \texttt{unknown}. These labels were used only for grouping; they were not included as model features.

\textit{Transfer experiments.} For each grouping variable and popularity indicator, we ran three experiments. 
We included a group only if it contained at least 30 popular and 30 unpopular models.
The within-group baseline trained and tested the Random Forest inside the same group using 10-fold cross-validation. One-versus-one transfer treated each group as a test target in turn: for a target group, we trained separate models on each other valid group and reported the mean AUC on the target group. Others-versus-one also treated each group as a target, but trained one model on the pooled data from all other valid groups before testing on the target group. For one-versus-one and others-versus-one transfer, we estimated test-set AUC with 1{,}000 bootstrap resamples~\cite{tibshirani1993introduction} and reported the mean and 95\% confidence interval. The within-group baseline measured prediction performance inside a group. The gap between within-group and one-versus-one AUC indicated how much performance was lost when training data came from another group, and the difference between one-versus-one and others-versus-one measured whether pooling multiple source groups improved prediction.

\subsection{Results}\label{subsec:rq3-result}

After applying the minimum-sample filter, we retained five domain groups: NLP, CV, Multimodal, Audio, and Other. We also retained five affiliation groups: \texttt{organization or individual}, \texttt{company}, \texttt{community}, \texttt{university}, and \texttt{non-profit}. The \texttt{classroom}, \texttt{government}, and \texttt{unknown} groups were excluded because their popular groups were too small. We report the per-group results in Table~\ref{tab:rq3-domain} and Table~\ref{tab:rq3-affil}.

\textit{Cross-domain transfer.} Table~\ref{tab:rq3-domain} reports the three AUC scores for each valid domain. The within-group AUC ranged from 0.811 (Other, downloads) to 0.891 (Computer Vision, downloads), and every within-group result exceeded the 0.7 reference threshold used in RQ2. One-versus-one transfer was lower than the within-group baseline for every domain. The average drop was 0.149 on downloads and 0.123 on likes. This shows that cross-domain prediction was harder than prediction within the same domain, even though AUC remained above random. The largest drop appeared for NLP under downloads, where AUC decreased from 0.861 within-group to 0.664 in one-versus-one transfer. Pooling all other domains improved over the average one-versus-one result for every domain, by 0.037 AUC on downloads and 0.050 on likes. This result suggests that cross-domain prediction benefits from seeing several source domains rather than only one.

\begin{table}[!htbp]
\centering
\small
\caption{Cross-domain transfer AUC, per domain, on the two popularity indicators. ``Within'' trains and tests inside the row domain with 10-fold cross-validation. ``1-vs-1'' treats the row domain as the test target and reports the mean AUC of models trained separately on each other valid domain. ``Others-vs-1'' trains one model on all other valid domains pooled together and tests it on the row domain. Sample sizes are shown in parentheses.}
\label{tab:rq3-domain}
\begin{tabular}{l c c c}
\hline
\textbf{Domain} & \textbf{Within} & \textbf{1-vs-1} & \textbf{Others-vs-1} \\
\hline
\multicolumn{4}{l}{\textit{by downloads}} \\
Audio (2,187)           & 0.814 & 0.684 & 0.690 \\
CV (7,853) & 0.891 & 0.753 & 0.776 \\
Multimodal (1,913)      & 0.844 & 0.712 & 0.761 \\
NLP (27,809)            & 0.861 & 0.664 & 0.713 \\
Other (20,276)          & 0.811 & 0.662 & 0.720 \\
\hline
\multicolumn{4}{l}{\textit{by likes}} \\
Audio (2,164)           & 0.824 & 0.735 & 0.781 \\
CV (7,521) & 0.852 & 0.716 & 0.756 \\
Multimodal (1,942)      & 0.832 & 0.713 & 0.769 \\
NLP (28,302)            & 0.848 & 0.708 & 0.754 \\
Other (20,103)          & 0.816 & 0.686 & 0.748 \\
\hline
\end{tabular}
\end{table}

\textit{Cross-affiliation transfer.} Table~\ref{tab:rq3-affil} reports the same three AUCs for the five valid affiliation groups. Within-group prediction was again strong. The highest within-group AUCs appeared for \texttt{company} models (0.881 on downloads and 0.867 on likes). One possible explanation is that company releases provide the repository-level cues used by our classifier more consistently, such as structured metadata, licensing, and complete model cards. The lowest within-group AUC appeared for \texttt{university} models under likes (0.765), suggesting that likes for academic releases may depend more on factors outside our repository-level features, such as paper visibility or author reputation.

One-versus-one transfer dropped more under affiliation grouping than under domain grouping: the average drop was 0.171 on downloads and 0.162 on likes, compared with 0.149 and 0.123 for domains. This difference suggests that contributor affiliation separated models by release practice more strongly than task domain did. Domains such as NLP, vision, audio, and multimodal describe what a model does, while affiliations also reflect who releases the model and under what institutional context. As a result, affiliation groups may differ more in metadata completeness, documentation style, and licensing practice, which are the types of features used by the classifier. The largest drop appeared for \texttt{community} under likes, where AUC decreased from 0.865 within-group to 0.677 in one-versus-one transfer. Pooling all other affiliations again improved over one-versus-one for every group, by an average of 0.033 AUC on downloads and 0.031 on likes.

\begin{table}[!htbp]
\centering
\small
\caption{Cross-affiliation transfer AUC, per affiliation, on the two popularity indicators. Columns are defined as in Table~\ref{tab:rq3-domain}: the row affiliation is always the evaluation target for ``1-vs-1'' and ``Others-vs-1''. Sample sizes are shown in parentheses.}
\label{tab:rq3-affil}
\begin{tabular}{l c c c}
\hline
\textbf{Affiliation} & \textbf{Within} & \textbf{1-vs-1} & \textbf{Others-vs-1} \\
\hline
\multicolumn{4}{l}{\textit{by downloads}} \\
community (5,353)                         & 0.868 & 0.681 & 0.728 \\
company (9,700)                           & 0.881 & 0.692 & 0.708 \\
non-profit (2,374)                        & 0.817 & 0.674 & 0.708 \\
organization or individual (40,168)       & 0.831 & 0.674 & 0.705 \\
university (2,403)                        & 0.839 & 0.664 & 0.701 \\
\hline
\multicolumn{4}{l}{\textit{by likes}} \\
community (5,298)                         & 0.865 & 0.677 & 0.716 \\
company (9,510)                           & 0.867 & 0.679 & 0.707 \\
non-profit (2,317)                        & 0.839 & 0.697 & 0.722 \\
organization or individual (40,472)       & 0.825 & 0.665 & 0.717 \\
university (2,402)                        & 0.765 & 0.635 & 0.645 \\
\hline
\end{tabular}
\end{table}

\begin{rqsummarybox}
\rqsummarylabel{3} The RQ2 Random Forest retained strong within-group performance across the five valid domains and five valid affiliations, with AUCs from 0.765 to 0.891. One-versus-one prediction was consistently lower than within-group prediction, dropping by 0.12--0.17 AUC on average, showing that performance was partly group-specific. The drop was larger across contributor affiliations than across domains. Training on all other groups together improved AUC by 0.031--0.050 over average one-versus-one prediction for every target group.
\end{rqsummarybox}

\section{Discussion}\label{sec:discussion}

This section examines the stability and interpretation of the results. We first report two robustness checks, then use sampled model cards to interpret the documentation findings, and finally provide practical recommendations for model practitioners.

\subsection{Robustness to High-Download Outliers}\label{subsec:disc-outliers}

A concern with the downloads indicator is that it includes both user-initiated retrievals and automated calls from scripts or applications. Prior work on software repository popularity has noted that engagement metrics such as GitHub stars are skewed and heavy-tailed~\cite{borges2016predicting}. Empirical SE guidelines also recommend checking whether conclusions are sensitive to extreme observations~\cite{kitchenham2017robust}. We therefore tested whether the main findings depended on a small number of extreme values by removing the top 1\% most-downloaded models (670 models, leaving 66{,}239 models) and rerunning the full analysis: RQ1 significance testing, RQ2 classifier comparison and permutation importance, and RQ3 within / 1-vs-1 / others-vs-1 transfer.

The main results remained largely unchanged after removing these high-download models. For RQ1, the count of FDR-significant features changed by at most two (downloads 28 to 27, likes 29 to 27), and no feature that reached at least small effect in the main analysis fell to negligible after cleaning. For RQ2, every classifier's AUC dropped by at most 0.017, and the model ranking was unchanged. Permutation-importance rankings on the best Random Forest correlated with the main analysis at Spearman $\rho = 0.97$ on downloads and $\rho = 0.99$ on likes, with the top-10 features overlapping 10/10 and 9/10. For RQ3, no within-group AUC shifted by more than 0.03, and only four of the 80 cross-group 1-vs-1 cells changed by more than 0.03, all with $|\Delta\text{AUC}| \leq 0.05$. The 0.12--0.17 within-to-1-vs-1 drop and the improvement from others-versus-one training also remained.

These results indicate that the main findings were not explained by a small number of high-download models. The slight reduction in classifier AUC (mean $-0.01$) is consistent with the cleaned data being slightly harder to classify after the most extreme popular examples were removed.

\subsection{Robustness to the Popular-Group Threshold}\label{subsec:disc-thresholds}

Our main scheme compared the top 10\% (popular) against the bottom 80\% (unpopular), with a 10\% gap buffer. This is one reasonable labelling choice, but not the only possible one. We therefore reran the analysis under three alternative schemes (5-10-85, 15-10-75, and 20-10-70) and compared them with the main 10-10-80 scheme.

Classifier AUC moved monotonically with the popular-group threshold: tightening the threshold to the top 5\% raised AUC by 0.024--0.040 because the popular group became more distinct, while loosening it to the top 15\% or 20\% lowered AUC by 0.012--0.030 because the popular group included more boundary cases. This reflects a change in task difficulty rather than a change in the main conclusions. The model ranking was preserved across all four thresholds, and Random Forest remained the best classifier in every setting. Permutation-importance rankings correlated with the main analysis at $\rho \geq 0.92$ under every threshold, and the top feature (\texttt{length-doc}) was unchanged. Cross-group transfer also followed the same structure: the within-to-1-vs-1 drop stayed in the 0.10--0.17 range, others-versus-one improved AUC by 0.03--0.05 over one-versus-one, and \texttt{university} remained the lowest within-group AUC on likes under every threshold.

\subsection{Qualitative Review of Sampled Model Cards}\label{subsec:disc-qualitative}

To interpret the Documentation features identified in RQ1 and RQ2, we sampled popular model cards from the main 10-10-80 data: 10 cards from the downloads-based popular group and 10 cards from the likes-based popular group. Two cards appeared in both samples, leaving 18 unique popular model cards. We then examined how the most important documentation features appeared in these sampled model cards.

The 18 sampled popular cards had a mean \texttt{length-doc} of 1{,}129 words and a median of 1{,}038.5 words, compared with medians of 451 words for the overall downloads-popular group and 483 words for the overall likes-popular group. Most sampled popular cards also contained the concrete elements highlighted by RQ1 and RQ2: 17 of 18 had at least one code block, 16 of 18 had at least one GitHub link, 16 of 18 had at least one arXiv entry, and 8 of 18 contained a BibTeX block. These observations provide concrete examples of why \texttt{length-doc}, \texttt{num-code-block}, \texttt{num-gh-links}, \texttt{num-arxiv-links}, and \texttt{has-bibtex} appear repeatedly in the quantitative results. They correspond to model-card content that helps a practitioner understand what the model is, try it, trace it to related work or code, and cite it.

When reading these cards, we also observed one documentation practice that is not directly captured by the 31 features: explicit intended-use, limitation, risk, or misuse sections. A keyword check found such content in 12 of the 18 sampled popular cards. For example, the \texttt{CompVis/stable-diffusion-v1-4} card\footnote{\url{https://huggingface.co/CompVis/stable-diffusion-v1-4}} contains sections on the intended use of the model with the Diffusers safety checker, misuse, out-of-scope use, limitations, and bias. Our current features capture the length of the card and the presence of links or code, but they do not distinguish a general long model card from one that explains limitations and responsible use. This suggests a possible direction for future feature extraction: model-card sections about intended use, limitations, and risks may deserve explicit features rather than being absorbed into \texttt{length-doc}.

\subsection{Practical Recommendations for Model Practitioners}\label{subsec:disc-recs}

Taken together, the quantitative results and qualitative review point to three practical recommendations for Hugging Face model practitioners. These recommendations are not causal claims. Rather, they describe practices that repeatedly appear in popular models and that are directly under practitioners' control when preparing a model repository.

\textit{Model-card content.} A model card should give enough information for a practitioner to decide whether and how to try the model. In the main analysis, popular model cards had a median \texttt{length-doc} of 451 words under downloads and 483 words under likes, compared with about 230--240 words for unpopular model cards. The strongest documentation features point to a concrete structure: include at least one runnable code block with imports and an example invocation (\texttt{num-code-block} and \texttt{num-inline-code}), link to the underlying paper or implementation when available (\texttt{num-arxiv-links} and \texttt{num-gh-links}), and provide a BibTeX block when the model is tied to a publication (\texttt{has-bibtex}). The qualitative review further suggests adding a short limitations or intended-use paragraph following the standard model-card template, even though this practice is not directly measured by the 31 features.

\textit{Model files and formats.} A model repository should make the model easy for Hugging Face tooling to load and inspect. In our results, \texttt{has-config} reaches small effect on downloads and appears among the top permutation-importance features, while \texttt{has-safetensors} also reaches small effect on downloads. The number of root-level files also appears in the RQ2 top-10 under both indicators, suggesting that the file layout itself carries information about how complete and inspectable a repository is. \texttt{model-size} is also highly ranked in RQ2, but it should not be read as a recommendation to make models larger. Instead, it shows that model scale is part of the context in which popularity is evaluated, and that repository files should make the model's size and loading requirements clear. For quantized models, practitioners can still release quantized models, but the model card should clearly state the quantization method, the source or base model, and the expected trade-off between efficiency and quality. This is especially important because quantized models are more concentrated in the unpopular group under likes, even though quantization remains important in RQ2.

\textit{Platform metadata.} Practitioners should fill the structured metadata fields used by the Hugging Face Hub, following the official model-card metadata instructions.\footnote{\url{https://huggingface.co/docs/hub/model-cards}} In particular, they should set a \texttt{pipeline\_tag}, declare the main implementation library through \texttt{library\_name}, and, when applicable, list one of the officially supported libraries. In RQ1, \texttt{has-pipeline}, \texttt{has-impl-lib}, and \texttt{has-supported-lib} all reach small effect on downloads, while \texttt{length-yaml} and several metadata-related features appear in the RQ2 top features. If the model supports interactive use, adding a Space or a \texttt{widgetData} block can also make the repository easier to try from the Hub. Together, these features suggest a practical checklist for making model metadata more complete and easier to use. Beyond individual practitioners, these findings also suggest opportunities for platform support. Model-sharing platforms could provide lightweight checks for missing runnable examples, citation information, model-card metadata, or limitation and intended-use sections. Such checks would not predict quality directly, but they could help practitioners make model cards more complete and easier to inspect.

\section{Threats to Validity}\label{sec:threats}

\textit{Construct validity.} We measured popularity using two count-based indicators provided by Hugging Face: downloads and likes. Neither indicator is a perfect measure of adoption or community value. Downloads can include automated calls from scripts or applications, while likes require a logged-in user to actively click the like button. For this reason, we did not merge the two indicators. We reported their Spearman correlation ($\rho = 0.36$), analyzed every research question separately for downloads and likes, and removed the top 1\% most-downloaded models in Section~\ref{subsec:disc-outliers} to check whether high-tail download counts affected the main conclusions. Our 31 features are extracted from model content and Hub metadata. This makes the measurements reproducible, but it also means that they miss model-card sections that require semantic interpretation, such as intended-use, misuse sections, limitation, or risk.

\textit{Internal validity.} Two design choices could affect the results: the 10-10-80 popularity scheme and the classifier-evaluation design. We addressed the first issue by rerunning the analysis under three alternative threshold schemes in Section~\ref{subsec:disc-thresholds}. For classifier comparison, we used 10-fold cross-validation repeated 10 times and AUC as the primary score; repeated runs produced the same model ranking. The permutation-importance rankings in RQ2 depend on the best-performing Random Forest. A different classifier, or a future model with higher predictive performance, could rank the features differently. The Mann--Whitney U and Chi-square tests used in RQ1 do not assume normally distributed feature values. All RQ3 cross-group AUCs include 1{,}000 bootstrap resamples to account for sampling variation. We also do not control for all possible confounders, such as contributor reputation beyond the affiliation label, because our features are defined at the model-repository level rather than at the contributor level.

\textit{External validity.} Our findings reflect the state of the Hugging Face Hub on the data collection date, 2026-02-28. Hugging Face is a single platform, and the practices we identify may not transfer directly to model registries with different community norms, such as ONNX Model Zoo or PyTorch Hub. Several features are platform-specific, including \texttt{has-pipeline}, \texttt{has-safetensors}, and \texttt{has-widgets}, so they would need to be redefined for another registry. Within Hugging Face, the results may also change as recommendation systems, default model-card templates, and community practices evolve. Finally, the qualitative review in Section~\ref{subsec:disc-qualitative} draws on 18 sampled popular model cards, so it should be read as an interpretation of the main documentation features rather than as a complete account of model-card practices.

\textit{Conclusion validity.} We used several checks to reduce the risk of over-interpreting the results. For RQ1, we applied the Benjamini--Hochberg false discovery rate procedure across all 31 feature tests under each indicator. For the overall pipeline, we reran the analysis after removing high-download outliers and under three alternative popular-group thresholds. These checks reduce the chance that the findings depend on one outlier treatment or one threshold scheme. The qualitative review has a narrower role: it is based on a small sample of popular model cards and is used only to interpret the documentation features identified by RQ1 and RQ2. Most importantly, our analyses are observational. They identify feature differences and predictive importance, but they do not establish causal effects between repository practices and popularity.

\section{Conclusion}\label{sec:conclusion}

We studied 66{,}909 Hugging Face models collected on 2026-02-28 and extracted 31 features across four dimensions: Documentation, Models, Platforms, and Techniques. We analyzed downloads and likes separately because they capture different forms of platform engagement.

Across both indicators, Documentation features provided the most consistent differences between the popular and unpopular groups. RQ1 identified 13 features on downloads and 8 features on likes that reached at least small feature-level effect. RQ2 showed that these differences also mattered in prediction: Random Forest reached AUC 0.861 on downloads and 0.848 on likes with top permutation-importance features such as \texttt{length-doc}, \texttt{length-yaml}, and \texttt{num-code-block}. RQ3 showed that prediction performance was partly group-specific. Within-group AUC ranged from 0.77 to 0.89, while one-versus-one transfer dropped by 0.12--0.17 AUC on average; training on all other groups improved AUC by 0.03--0.05 over one-versus-one transfer.

The main findings remained stable after two robustness checks: removing the top 1\% most-downloaded models and rerunning the pipeline under three alternative popular-group thresholds. A manual review of 18 sampled popular model cards further showed how the main documentation features appear in real model cards: the sampled cards were longer than the overall popular-group median, usually included code blocks and links to papers or code, and sometimes included BibTeX blocks. The review also pointed to one practice not captured by our 31 features: explicit limitations and intended-use sections.

These findings support a small set of recommendations for Hugging Face model practitioners. Model cards should describe a complete usage flow, include runnable code examples, link to the underlying paper or implementation when available, and provide citation information when appropriate. Model repositories should include loadable configuration files and SafeTensors weights when possible, and practitioners should fill structured metadata fields such as \texttt{pipeline\_tag}, \texttt{library\_name}, supported-library declarations, and Spaces or widgets when applicable. Future work could extract limitations and intended-use sections explicitly, replicate the analysis on registries beyond Hugging Face, and study how these popularity-related features evolve as the platform's recommendation system and default model-card templates change.

\section*{Data Availability}
The replication package is publicly available at:
\url{https://github.com/YinanWusoymilk/What-Makes-HF-Models-Popular}.

\bibliography{references}

\end{document}